# WAFERSCALE ELECTROSTATIC QUADRUPOLE ARRAY FOR MULTIPLE ION BEAM MANIPULATION

K.B. Vinayakumar[1*], A. Persaud[2], Q. Ji[2], P. Seidl[2], T. Schenkel[2] and A. Lal[1]
[1]*Sonic*MEMS Laboratory, Cornell University, Ithaca, NY 14853, USA
[2]Lawrence Berkeley National Laboratory, Berkeley, CA 94720, USA

## ABSTRACT

We report on the first through-wafer silicon-based Electrostatic Quadrupole Array (ESQA) to focus high energy ion beams. This device is a key enabler for a wafer-based accelerator architecture that lends itself to orders-of-magnitude reduction in cost, volume and weight of charged particle accelerators. ESQs are a key building block in developing compact Multiple Electrostatic Quadrupole Array Linear Accelerator (MEQALAC) [1]. In a MEQALAC electrostatic forces are used to focus ions, and electrostatic field scaling permits high beam current densities by decreasing the beam aperture size for a given peak electric field set by breakdown limitations. Using multiple parallel beams, each totaling to an area A, can result in higher total beam current compared to a single aperture beam of the same area. Smaller dimensions also allow for higher focusing electric field gradients and therefore higher average beam current density. Here we demonstrate that Deep Reactive Ion Etching (DRIE) micromachined pillar electrodes, electrically isolated by silicon-nitride thin films enable higher performance ESQA with waferscale scalability. The fabricated ESQA are able to hold up to 1 kV in air. A 3×3 array of 12 keV argon ion beams are focused in a wafer accelerator unit cell to pave the way for multiple wafer accelerator.

## INTRODUCTION

Particle accelerators produce high intensity charged particle beams for basic science (atomic, nuclear, particle physics) as well as applications such as ion-propulsion, medical treatment, surface treatments, neutron and x-ray generation, etc [2]. The size, weight and power (SWAP) of existing accelerators precludes their wider use in portable applications, and the cost precludes their availability in applications such as medical proton beam therapies in remote areas. We aim to reduce the SWAP and cost by developing an accelerator that using a sequence of easily replaceable wafers to accelerate and focus a densely packed array of parallel ion beams. Each wafer can be mass-manufactured using lithographic patterning and micromachining methods.

Figure 1 shows the MEQALAC concept. The key components required to develop a MEQALAC are an ion source, an ion extraction unit, focusing elements and accelerating stages [1-4]. Focusing elements are used to keep the beam in-line with the main direction of motion as self-repulsion of charged particles in the beam increases the beam diameter. The accelerating stages are used to increase the beam power while it propagates along the beam line. To improve the accelerator performance, different approaches to design focusing and accelerating components have been explored [2, 5, 6].

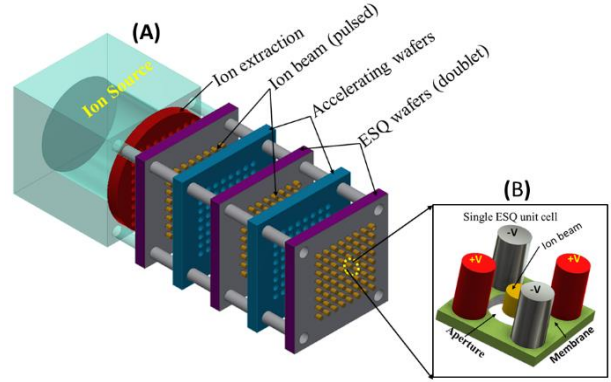

*Figure 1: (A) 3D schematic of multi-beam accelerator with ion source, focusing and accelerating wafers (B) Single ESQ unit cell (conducting pillars) on insulating membrane with beam in the aperture.*

In practice, ion beams will have a small velocity perpendicular to the direction of motion. This can be due to space charge forces, mechanical imperfections in the accelerator structure or thermal motion of particles at the source [2]. Electric or magnetic fields can be used to manipulate the charged particles and, for example, refocus them along the beam axis [5-7]. Previous work demonstrated the use of electrostatic fields on a wafer-scale device to focus and bend a single ion beam along a path on the wafer [5-8], while in the work presented here through-wafer focusing and acceleration is used.

In a previous publication we showed that a printed-circuit-board based (PCB) ESQA can be used to focus and accelerate argon ion beams [9, 10]. However, the laser microfabrication and side-wall metal deposition processes are not suitable for mass fabrication. The sidewall roughness due to laser processing will also limit the achievable electric field. To improve the ESQA performance, this paper will discuss the design, fabrication and testing of a 3×3 silicon ESQA. The silicon ESQA was fabricated using standard lithography-based microfabrication techniques. This allows the reduction of the ESQ unit cell dimension thereby increasing the effective beam current densities. The fabricated device can hold voltage up to 1 kV in air over the opposing polarity electrodes separated by 250 μm. To observe beam focusing and defocusing effects, tests were carried out using an argon beam extracted from a multi-cusp plasma ion source.

## DEVICE DESIGN

An ESQ consists of four electrodes spaced symmetrically around the beam axis (Figure 2) where voltages of alternating polarity create an electric field that focuses ions in one direction and defocuses the ions in the other direction.

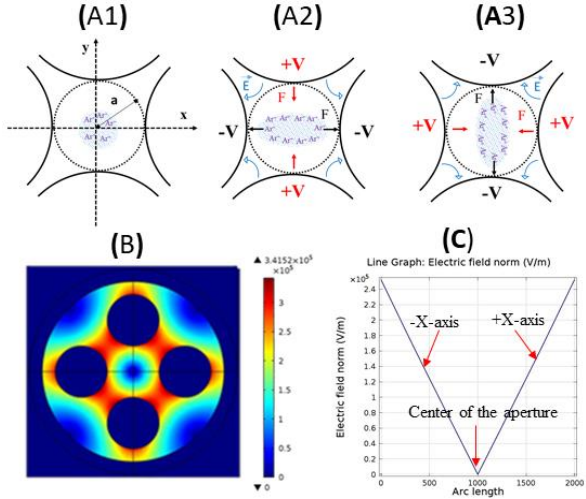

*Figure 2: (A) Focusing effect on a positive ion-beam A1-beam with no potential applied on the ESQ A2-beam focusing in one direction according to the potential shown A3-focusing direction changes by changing the polarity on the ESQ. (B) COMSOL simulation of the electric field distribution (circular electrode with 2mm dimeter with 2mm beam aperture at the center, V=±125V). (C) Shows the electric field distribution in the x-axis, where electric field at the center of the beam aperture is zero.*

Combining two ESQs into a doublet results in an overall focusing effect in both lateral directions. The focal length of the lens scales inversely with the electric field gap favoring the scaling of the ESQ electrode diameter and spacing to smaller dimensions. Since ESQ focuses in one direction and defocuses in another the value of the lens can be questioned. However, if a ESQ lens is followed by another, one with focusing orthogonal to the focusing of the next lens, ions entering the lens in one direction are deflected towards the corresponding for the lens. When the focused/defocused ions arrive at the second lens, they are closer to the axis and the outward forces and defocusing is reduced, leading to a net focusing.

In the present work, a highly doped silicon wafer was used to fabricate ESQA electrodes. Highly doped silicon high aspect ratio conductive pillar structures were fabricated on an oxide-nitride membrane. Membrane stress was optimized by adjusting the oxide and nitride thickness ratio (1 µm silicon oxide and 2 µm silicon nitride) (silicon nitride with tensile stress of ~200Mpa and silicon oxide with compressive stress of ~400Mpa). To increase the breakdown voltage, high aspect ratio structures were optimally spaced (~250 µm) and placed on a thick oxide-nitride membrane. The electrodes were wire bonded for two-polarity voltage applications to the pillar arrays.

Figure 2(A1) shows the center beam with no voltages applied on the ESQ, Figure 2(A2) and Figure 2(A3) illustrate the beam focusing and de-focusing effect on positive ions for different applied polarities. The electrostatic force on a positive ion will be repulsive for a positive potential and attractive for a negative potential. Resulting in an elliptical beam shape for an initial round beam. Equations for the potential distribution in the ESQA is given by:

$$V(x,y) = \left(\frac{V_0}{a^2}\right)(y^2 - x^2) \quad \text{---------(1)}$$

Where, $V(x,y)$ is the electric potential, $a$ is the minimum distance between the axis to the electrode, $V_0$ is the electric potential on the electrode surface.

The two dimensional field for the electrostatic quadrupole is given by below equations:

$$E_y = E_0\left(\frac{y}{a}\right) \quad \text{--------(2)}$$

$$E_x = -E_0\left(\frac{x}{a}\right) \quad \text{--------(3)}$$

$$E_0 = \frac{2V_0}{a} \quad \text{--------(4)}$$

Where, $E_0$ is the electric field on the electrode surface, Force acting on the ion beam is given by:

$$\vec{F} = q\vec{E} + \cancel{q\vec{v} \times \vec{B}} \quad \text{--------(5)}$$

Where, $F$ is the force acting on the charged particles, $q$ is the charge and $B$ is the magnetic field (which is not considered in our present study).

A COMSOL simulation in Figure 2(B) illustrates the electric field distribution of an ESQ. Figure 2(C) shows electric field distribution in an ESQ. The electric field at the center of the quadrupole aperture is zero and increases (or decreases) linearly with distance from the center.

## DEVICE FABRICATION

Silicon ESQAs have been fabricated using conventional silicon micro-fabrication techniques. Figure 3 shows the fabrication process flow. A highly doped silicon wafer (100 mm 4", resistivity 0.005-.020 Ω cm, thickness 490-510 µm) was used for the process. The doped silicon wafer was thermally oxidized (1 µm) and silicon nitride (2 µm) was deposited for electrical isolation and as a support layer. A first lithography step was carried out to pattern the beam aperture and metallization area. After patterning, Reactive Ion Etching (RIE) was used to etch the nitride and oxide layers to reach the silicon. Then a nickel was evaporated (100 nm) to form an electrical contact with the doped silicon pillars. After the metal deposition, a Plasma-enhanced chemical vapor deposition (PECVD) oxide was deposited (2 µm) this is required to isolate the metal from the plasma of the DRIE process. A second lithography step was carried out to pattern the electrode pillars and isolation region. After patterning, RIE was used to etch the nitride and oxide layers to reach the silicon. A DRIE process was used to etch silicon and form electrode pillars (with side wall roughness of ~30nm) [11]. An oxide isolation layer was deposited to avoid wire bond contact with the underlying metal layer (this in result increases the breakdown potential). Then, the wire-bond was carried out to interconnect the electrodes of same polarity. The wirebonding height limits the stacking density of different ESQ wafers. By stacking two ESQ pillar wafers it is possible to eliminate the wirebonds, and this device fabrication is under investigation.

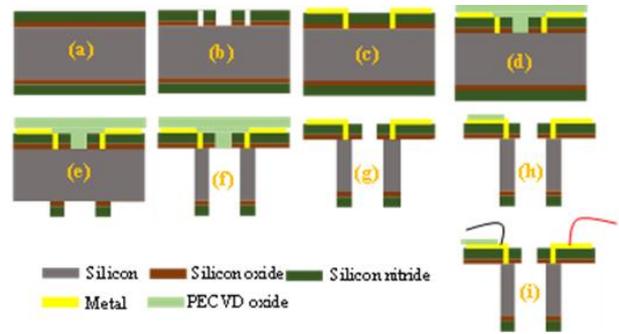

*Figure 3: (a) Low Pressure Chemical Vapor Deposition (LPCVD) nitride deposition (2000 nm) and oxide growth (1000 nm) on highly doped silicon wafer (b) Patterned oxide and nitride for metal deposition (c) Metal (nickel) e-beam evaporation (100 nm) (d) PECVD oxide on front side-stop layer for DRIE (e) Patterned back side oxide and nitride (f) DRIE from back side (g) Removed the front side PECVD oxide for aperture (h) Oxide to isolate wire bond crossing (i) Wire bond to interconnect +ve and –ve silicon electrodes.*

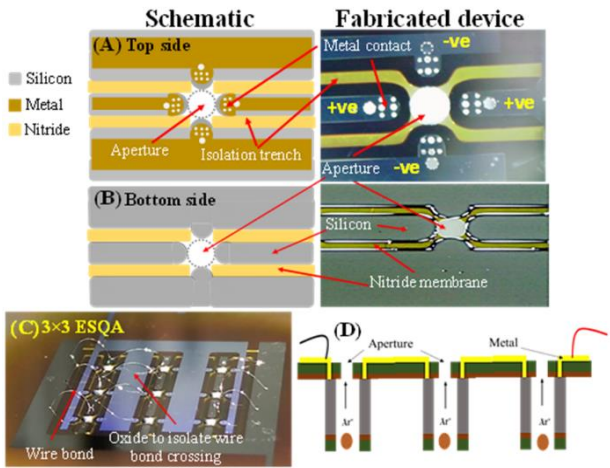

*Figure 4: (A) Top view of the ESQA with metal contact and separation of ~250 micron between the oppositely charged pillars. (B) Bottom view of the ESQA with 1 mm aperture diameter and standing electrode pillars on oxide/nitride membrane (C) 3×3 ESQA with 5 mm spacing, array shows the wire bond crossing on nitride membrane (to increase the breakdown potential) (D) Cross section of 3×3 ESQA with ion beam moving in aperture.*

The fabricated silicon ESQA is shown in Figure 4. A top view of the fabricated wafer in Figure 4(A) shows the metal routing connecting the electrode pillars. The bright circular dots on the electrode pillar depicts the metal contact with silicon and the light blue color of the metal indicate the metal routing on the nitride membrane. The yellow region in the fabricated device (Figures 4(A) and 4(B)) shows the oxide-nitride layer used to isolate the opposite polarity electrodes. The bottom side of the wafer shows the hyperbolic shaped electrode pillars around the beam aperture. The fabricated hyperbolic electrode pillars are perpendicular to each other to form the quadrupole structure. As mentioned before, this results in the typical linear varying electric field at the center of the beam aperture when all four electrode are biased at the same potential using alternating polarities. Figure 4(C) shows the fabricated ESQA with an array of 3×3 beam apertures, with a diameter of 1 mm for each aperture. The minimum spacing between electrode pillars of opposite polarities is ~250 µm, and the shortest path for breakdown either along the nitride support or the vacuum gap.

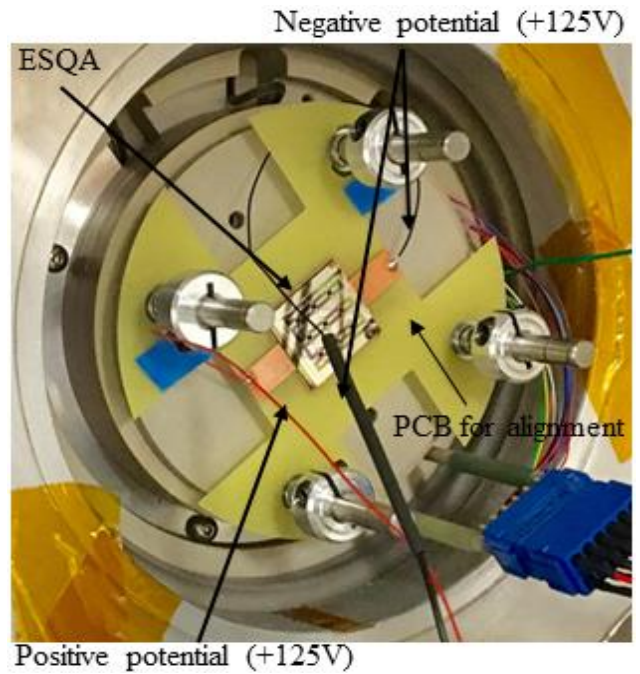

*Figure 5: Photo of the experimental configuration in the vacuum chamber, showing the ESQA immediately downstream of the injector and matching section.*

## BEAM TESTING AND RESULTS

Singly ionized argon ions were injected into the ESQA from a multi-cusp plasma ion source using a multi-grid extraction setup, as shown in Figure 5 and 6. The injection energy was 12 keV, and a 3×3 array aperture extraction array injected the beamlets into a six quadrupole matching section and subsequently into the ESQA. A scintillator (RP 400 plastic scintillator) was mounted downstream of the ESQA. A fast imaging intensifying camera (Princeton Instruments) measured the beam intensity pattern several centimeters downstream of the exit of the ESQA for different focusing voltages applied to the quadrupole electrodes. Focusing and defocusing of the beam was observed as predicted when electrode polarities (±125 V) are reversed, and a round beam was observed at zero bias (Figure 7).

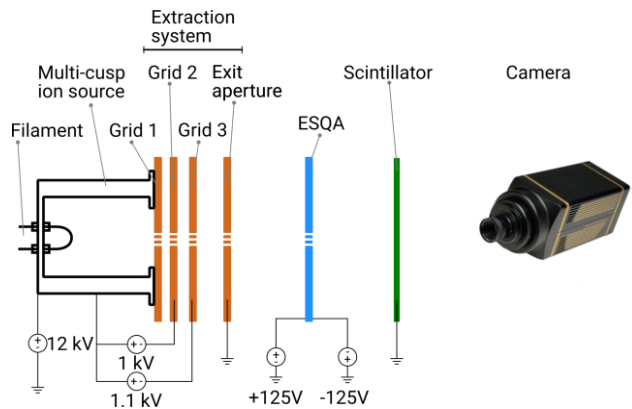

*Figure 6: Schematic of the experimental setup used to test the beam focusing (average beam current ~50 µA). A scintillator measures the transverse beam distribution, shown below.*

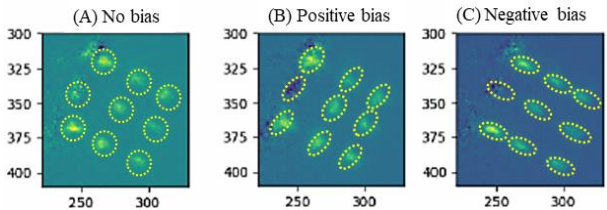

*Figure 7: ESQA focusing effect for ±125 V (left image) and ∓125 V (center) bias recorded using a scintillator and a CCD camera. As expected focusing in one (diagonal) direction and defocusing in the orthogonal direction is observed, with the pattern inverted with the change in polarity. Common background light from the filament of the source is subtracted.*

A ±120 µA leakage current through residual silicon and FR4 PC board was measured, implying R ≈ 2 MΩ between opposing polarities.

In a second beam test the focusing effect was not as pronounced, due perhaps to one or more failed wire bonds or other electrical connections which are quite delicate. Another possibility is that over time tracking deposits led to increased leakage current and diminished voltage holding on the pillar electrodes. Future designs will implement more robust connections and more leakage current diagnostics.

## SUMMARY AND FUTURE WORK

In the present work we demonstrated beam focusing with pairs of electrostatic quadrupoles based on microfabricated silicon structures in a 3×3 ESQA. The fabricated device is able to withstand up to 1 kV in air. We aim to study the maximum fields that can be achieved with the ESQA for probing the maximum limits of the effectiveness of one ESQA wafer. In the current work, it is evident that the average beam current can be increased by increasing the number of beam apertures. Along with the ESQA we are working on integrating on-board resonators as accelerating stages to form a complete compact accelerator. In the near future, we aim to improve the performance of the ESQA by stacking two ESQA wafers to eliminate the need for wirebonds. This process flow requires the development of bonding of two DRIE etched wafers.

## ACKNOWLEDGEMENTS

This work was supported by the Office of Science of the U.S. Department of Energy through the ARPA-E ALPHA program under Contract No. DE-AC0205CH11231. Device fabrication was carried out at the Cornell Nano Fabrication (CNF) facility, a member of the National Coordinate Science Foundation (NNCI) network, supported by the National Science Foundation (Grant No. ECCS-1542081).

## CONTACT

*K B Vinayakumar, tel: +1-607-3791720;
vk256@cornell.edu, vinayjgi@gmail.com